\documentclass{jetpl}
\twocolumn
\lat

\title{Electron energy level statistics in graphene quantum dots}

\rtitle{Electron energy level statistics in graphene quantum dots}

\sodtitle{Electron energy level statistics in graphene quantum dots}

\author{H.\, De Raedt$^*$ and
M.\, I.\, Katsnelson$^+$\/\thanks{e-mail: M.Katsnelson@science.ru.nl}}

\rauthor{H.\, De Raedt and M.\, I.\, Katsnelson}

\sodauthor{De Raedt, Katsnelson}

\address{$^*$Department of Applied Physics,
Zernike Institute for Advanced Materials \\University of
Groningen, Nijenborgh 4, NL-9747 AG Groningen, The Netherlands\\~\\
$^+$Institute of Molecules and Materials, Radboud
University of Nijmegen \\ NL-6525 ED Nijmegen, The Netherlands}

\dates{12 August 2008}{12 September 2008}

\abstract{Motivated by recent experimental observations of size quantization
of electron energy levels in graphene quantum dots
\cite{ponomarenko} we investigate the level statistics in the
simplest tight-binding model for different dot shapes by computer
simulation. The results are in a reasonable agreement with the
experiment which confirms qualitatively interpretation of observed
level statistics in terms of ``Dirac billiards'' without taking
into account many-body effects. It is shown that edge effects are
in general sufficient to produce the observed level distribution
and that even strong bulk disorder does not change the results
drastically.}


\begin{document}

\maketitle

Graphene attracts enormous attention now, due to its interest both
for fundamental physics, such as opportunities to simulate in
condensed matter experiments subtle quantum relativistic effects
and for potential applications, as a planar, high-mobility
material for ``post-silicon'' electronics \cite{r1,r2,r3,r4,r5}.
Graphene-based nanodevices are subjects of especial interest.
Recently, size quantization effects have been observed in graphene
nanoribbons \cite{han} and quantum dots (QD) \cite{ponomarenko}.
It was demonstrated that for QD smaller than 100 nm the electron
energy spectrum is essentially irregular demonstrating a
``chaotic'' behavior. The latter can be discussed in terms of
random matrix theory for a single-particle problem
\cite{random1,random2,random3}. However, in general, because of
interplay of size quantization and Coulomb blockade correlation
effects can be important for QD \cite{shankar}; moreover, in some
limiting cases the ``chaotic'' energy spectrum can be described
even purely classically, in terms of {\it only} Coulomb energies
\cite{shklovskii}. Therefore, to understand properly the
experimental data \cite{ponomarenko} some theoretical efforts are
necessary.

In this Letter we present the results of straightforward computer
simulations of level statistics in graphene QD, adopting the simplest
one-electron picture, up to 16000 sites (which is comparable to
the sizes of smallest QD investigated experimentally).
It turns out that already this approach does allow to reproduce, in a
semi-quantitative way, the observed energy distribution.

\begin{figure}[t]
\begin{center}
\includegraphics[width=8cm]{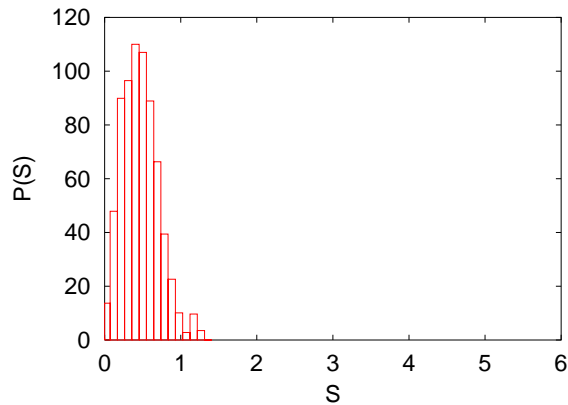}
\end{center}
\caption{Fig.~1 (Color online) Level-spacing distribution $P(S)$ as
obtained from experiment~\cite{ponomarenko} for the case of 40 nm
QD (The raw experimental data are by courtesy of K. Novoselov and
A. Geim). } \label{fig1}
\end{figure}

As the model Hamiltonian, we take the simplest nearest-neighbor
tight-binding model on a hexagonal lattice as introduced in
Ref.~\cite{WALL47}, enclosed in some geometrical shape, such
as a circle, triangle, etc.
The Hamiltonian reads
\begin{eqnarray}
H&=&-t\sum_{\langle i,j\rangle} \sum_{\sigma=\uparrow,\downarrow}
\left(a^{\dagger}_{i,\sigma}b^{\phantom{\dagger}}_{j,\sigma}
+
b^{\dagger}_{j,\sigma}a^{\phantom{\dagger}}_{i,\sigma}
\right)
\nonumber \\
&&+\sum_{i\in{\cal B}}\sum_{\sigma=\uparrow,\downarrow} V^{\phantom{\dagger}}_i a^{\dagger}_{i,\sigma}a^{\phantom{\dagger}}_{i,\sigma}
+\sum_{j\in{\cal B}} \sum_{\sigma=\uparrow,\downarrow}V^{\phantom{\dagger}}_j b^{\dagger}_{j,\sigma}b^{\phantom{\dagger}}_{j,\sigma}
\nonumber \\
&&+\sum_{i}\sum_{\sigma=\uparrow,\downarrow} v^{\phantom{\dagger}}_i a^{\dagger}_{i,\sigma}a^{\phantom{\dagger}}_{i,\sigma}
+\sum_{j} \sum_{\sigma=\uparrow,\downarrow}v^{\phantom{\dagger}}_j b^{\dagger}_{j,\sigma}b^{\phantom{\dagger}}_{j,\sigma}
,
\label{HAM0}
\end{eqnarray}
where $t$ is the nearest-neighbor hopping energy,
$a^{\dagger}_{i,\sigma}$ ($a^{\phantom{\dagger}}_{i,\sigma}$)
creates (annihilates) an electron with spin $\sigma$ on one of the
sub-lattices of the hexagonal lattice and
$b^{\dagger}_{j,\sigma}$ ($b^{\phantom{\dagger}}_{j,\sigma}$)
creates (annihilates) an electron with spin $\sigma$ on the
other sub-lattice of the hexagonal lattice.
In the expression for the hopping term, the indices $i$ and $j$ run over
all nearest neighbors only.

Following Ref.~\cite{akhmerov}, we also
consider the effects of a staggered on-site potentials $V_i$ and $V_j$ that
alternate as we move along the boundary ${\cal B}$ of the lattice; as shown in Ref.~\cite{akhmerov},
the continuum limit with a very large $V_i$ and $V_j$ corresponds to the
``infinite mass'' boundary condition \cite{random1}. The physical
origin of this potential can be related to, for instance, the magnetic
moments at the zigzag edges \cite{m1,m2}. Of course, in a
simulation we can easily study the effect of $V_i$ by considering the
cases $V_i=V_j=0$ and $V_i\not=0$ and $V_j\not=0$.
Optionally, to study the effect of disorder, we use uniform pseudo-random numbers to choose the
hopping integrals from the interval $[t-\delta,t+\delta]$ and/or
we add on-site potentials $v_i$ and $v_j$ in the range $[-v,v]$.

The geometrical shape ``cuts out'' a
piece of the hexagonal lattice such that armchair and
zigzag parts of the boundary appear (for a general discussion of
boundary conditions for the tight-binding model of graphene, see
Ref.~\cite{akhmerov}).
Furthermore, we study the effect of disorder that results from removing hexagons
from the regular hexagonal layer.

The eigenvalues of Hamiltonian Eq.~(\ref{HAM0}) are obtained by
exact numerical diagonalization. As we are
interested in the part of the spectrum that, in the continuum
limit corresponds to the spectrum of the Dirac equation, we limit
the search for eigenvalues to the interval $[-0.4 t,0.4 t]$. In
our numerical work, we adopt units such that $\hbar=1$
and we set the hopping integral $t=1$.

For reference, in Fig.~\ref{fig1}, we show the experimental
results of the level spacing distribution for a graphene dot of 40
nm diameter~\cite{ponomarenko}. The dimensionless level spacing
$S$ is defined as the energy difference $\Delta E_i=E_i-E_{i-1}$
between successive levels, divided by the average $\langle\Delta
E_i\rangle$ of the energy differences between successive levels.
The number $P(S)$ gives the number of energy difference for which
$S-\Delta/2<\Delta E_i/\langle\Delta E_i\rangle \le S+\Delta/2$,
where $\Delta$ is the bin size of the histogram.
The experimental data for the distribution have been
shifted by $0.02$ V to remove an ambiguity in the definition
of $S=0$.

\begin{figure}[t]
\begin{center}
\includegraphics[width=8cm]{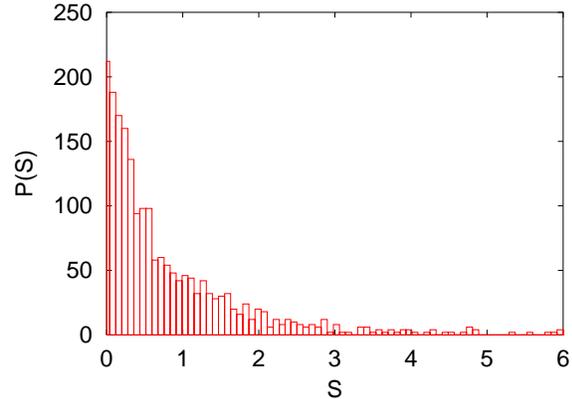}
\end{center}
\caption{Fig.~2 (Color online) Level-spacing distribution $P(S)$ for a
hexagonal lattice with periodic boundary conditions, using 170
different values for both $k_x$ and $k_y$.
}
\label{fig2}
\end{figure}

For comparison, Fig.~\ref{fig2} shows the level distribution for a
hexagonal lattice with periodic boundary conditions, a lattice
that has not been cut-out using some geometrical shape. It is
clear that its spectrum does not resemble the one observed
experimentally.

Next, we consider the case of a circular dot. In the continuum
approximation with the infinite mass boundary conditions
\cite{random1} this case is special, with separable variables and
a regular energy spectrum. In reality, although we use a circle to
cut out from the infinite lattice, the boundary of this lattice is
irregular: There are short and long pieces of the armchair and
zigzag boundaries. This kind of irregularities destroy the
circular symmetry of the the geometrical shape completely.

\begin{figure}[ht]
\begin{center}
\includegraphics[width=8cm]{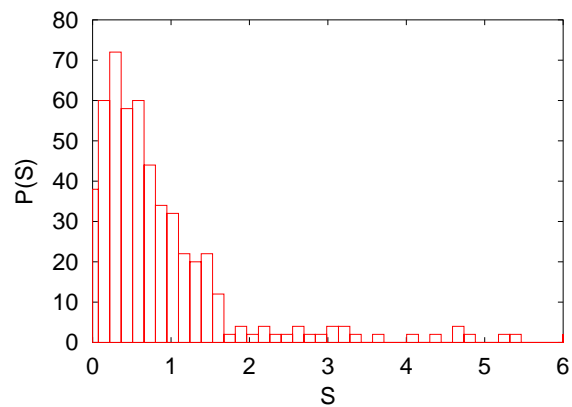}
\end{center}
\caption{Fig.~3 (Color online) Level-spacing distribution $P(S)$ for a
hexagonal lattice bounded by a circle. The number of hexagons
inside the circle is 8202. } \label{fig3}
\end{figure}

\begin{figure}[ht]
\begin{center}
\includegraphics[width=8cm]{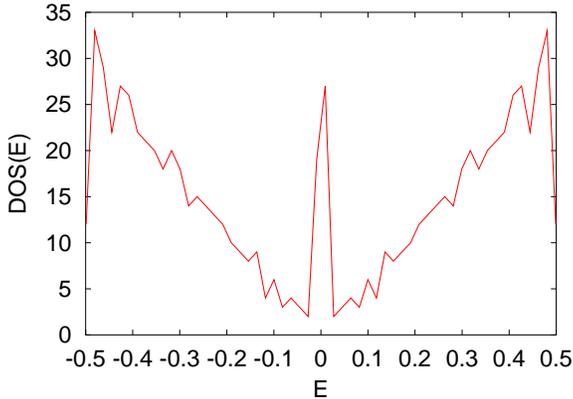}
\end{center}
\caption{Fig.~4 (Color online)
Density of states DOS$(E)$  as a function of the energy $E$
for a hexagonal lattice confined to a circle.
The number of hexagons inside the circle is 8202.
}
\label{fig4}
\end{figure}

\begin{figure}[ht]
\begin{center}
\includegraphics[width=8cm]{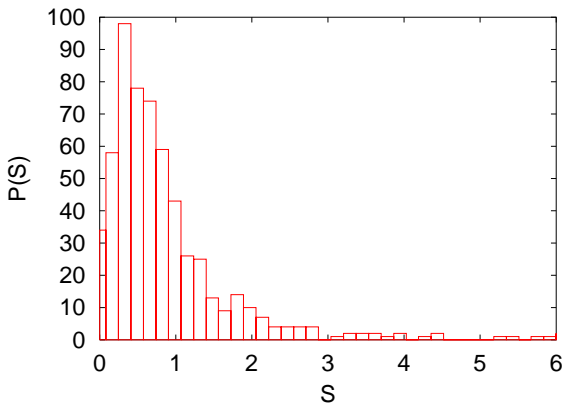}
\end{center}
\caption{Fig.~5 (Color online) Same as Fig.~\ref{fig3} except that the
staggered potential at the boundary edges $V=100$, the hopping
integrals fluctuate by maximum $20\%$ ($\delta=0.2$) and the
on-site potentials fluctuate in the range $[-0.2,0.2]$ The number
of hexagons inside the circle is 8484. } \label{fig5}
\end{figure}

\begin{figure}[ht]
\begin{center}
\includegraphics[width=8cm]{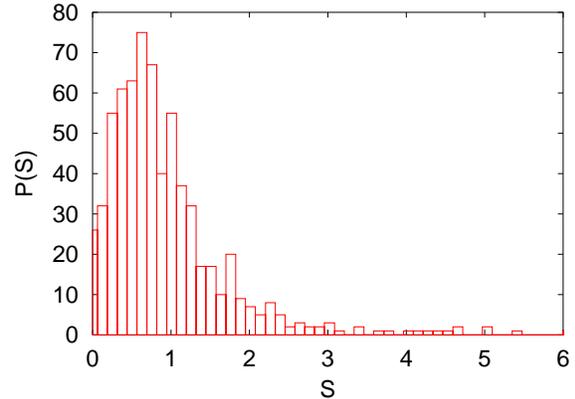}
\end{center}
\caption{Fig.~6 (Color online)
Same as Fig.~\ref{fig3} except that the potential at the boundary
edges $V=10$, about $20\%$ of the hexagons have been removed,
the hopping integrals fluctuate by maximum $20\%$ ($\delta=0.2$)
and the on-site potentials fluctuate in the range $[-0.2,0.2]$
The number of hexagons inside the circle is 6540.
}
\label{fig6}
\end{figure}

In Figs.~\ref{fig3} and \ref{fig4}, we depict the results for
$P(S)$ and the density of states DOS$(E)$ for the case of the
perfect hexagonal lattice, bounded by a circle. Although there is
a clear background linear dependence of the DOS$(E)$ on $E$, it is
also clear that there are fluctuations due to size quantization.
In the DOS, the peak around zero energy is due to the existence
of the zigzag edge states,
as expected for a generic boundary \cite{akhmerov}.
Comparing Fig.~\ref{fig3} with the experimental result
Fig.~\ref{fig1}, we conclude that there is little resemblance.

In contrast, by introducing various forms of disorder to the
same system, we find semi-quantitative agreement,
as shown in Figs.~\ref{fig5} and \ref{fig6}.
Our numerical experiments (not all results shown) suggest
that the presence of an alternating boundary potential
can change the qualitative features of $P(S)$ significantly.
Including various forms of disorder ($\delta\not=0$,
$v\not=0$ and remove some hexagons), the
level distribution $P(S)$ (see Fig.~\ref{fig6})
looks similar to the experimental result (see Fig.~\ref{fig1}).
Of course, we cannot expect quantitative agreement:
The number of items in the experimental data
is about 60 while in the numerical simulations
there are about 600 eigenvalues that contribute to $P(S)$.
This may explain why the simulation results
for $P(S)$ show a more extended tail than the experimental
result for $P(S)$.

\begin{figure}[t]
\begin{center}
\includegraphics[width=8cm]{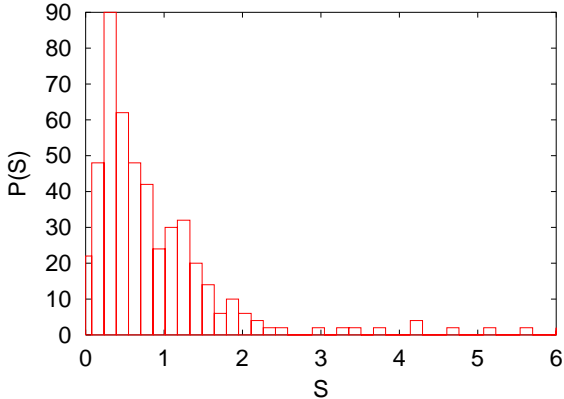}
\end{center}
\caption{Fig.~7 (Color online)
Level-spacing distribution $P(S)$ as a function of the level spacing $S$
for a hexagonal lattice bounded by a quarter of
a lemon-shaped billiard for $V=\delta=v=0$.
Parameter of the lemon: $ d=3.1$ ($d=1$ is half of a triangle, $d=2$ is a half of parabola).
The number of hexagons inside the billiard is 7861.
}
\label{fig7}
\end{figure}

\begin{figure}[t]
\begin{center}
\includegraphics[width=8cm]{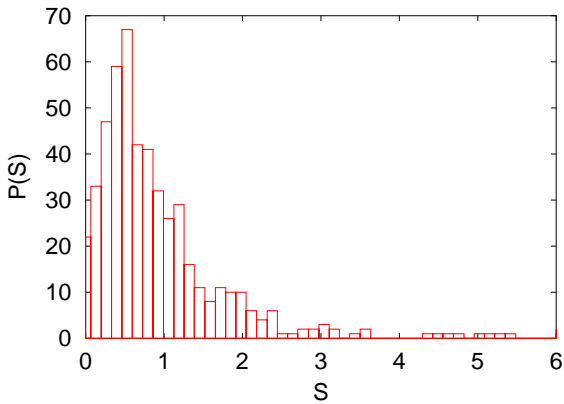}
\end{center}
\caption{Fig.~8 (Color online)
Same as Fig.~\ref{fig7} except that $v$ is chosen at random from $[-0.2,0.2]$.
}
\label{fig8}
\end{figure}

Finally, to study the effect of the shape of the boundary on the
level spacing distribution, we have calculates $P(S)$ for various
lemon-shaped billiards (results not shown) defined by
$y=\pm(1-|x|^d)$ for $x\in[-1,1]$~\cite{LOPA99}. As a function of
the shape parameters $d>0$, these billiards look like a square
($d=1,\infty$) or two parabola ($d=2$) or have some intermediate
lemon-like shape. It has been shown that, as a function of $d$,
the these billiard, classical as well as quantum mechanically,
exhibit regular and chaotic behavior~\cite{LOPA99}. If
Figs.~\ref{fig7} and \ref{fig8}, we present some typical results
for a quarter lemon with $d=3.1$. Our motivation for presenting
the results of a quarter lemon instead of the complete lemon is to
show a case in which there are long stretches of armchair and
zigzag boundaries (the edges along the $x$ and $y$ axis) and an
irregular boundary (the curved edge). Although on purpose, we did
not include the alternating potential at the boundary sites, the
disorder resulting from the irregular shape together with the
fluctuating on-site potential $v$ seem to be sufficient to observe
a $P(S)$ that is similar to the distribution observed
experimentally. Comparing Figs. 7 and 8 one can see that the level
statistics is not too sensitive to the bulk disorder. There is
only randomness due to boundaries themselves in one case (Fig. 7)
and a random potential $v$ is introduced, additionally, in the
other one (Fig. 8), but the results look very similar.

To conclude, it seems that disorder due to randomness of the
edges is, in principle, enough to reproduce the experimentally
observed level distribution which makes the term ``chaotic Dirac
billiard'' quite reasonable. One may need the local on-site
disorder and some disorder in the hopping integrals to get
semi-qualitative agreement with experiment but, on the other hand,
the simulated systems do not have the same shape and
are not as large as the experimental ones so a
complete quantitative agreement is, anyway, hard to expect.

It is worth mentioning that the continuum approximation may be a
bit dangerous when discussing the level statistics in graphene
QDs. In our simulations, we do not see any essential differences between ``regular''
(circular) and ``irregular'' (lemon-shaped) billiards. 
Even for tens of thousand sites the edge is essentially irregular with
staggered armchair and zigzag pieces. Therefore, even for circular
quantum dot we may have a ``chaotic'' energy level distribution.

We are thankful to Andre Geim and Kostya Novoselov for helpful
discussions and for providing us with the original experimental data used
in Fig.~\ref{fig1} and to Boris Shklovskii for illuminating discussions.
The work was supported by the Stichting voor Fundamenteel
Onderzoek der Materie (FOM) (the Netherlands)

\end{document}